\documentclass[twocolumn,prl,showpacs]{revtex4}

\usepackage{dcolumn}
\usepackage{bm}

\usepackage{graphicx}

\usepackage{amssymb,amsfonts}

\begin{document}



\title{Subharmonic gap structure in superconducting scanning tunneling microscope junctions}


\author{O.~Naaman}
\author{R.~C.~Dynes}
\affiliation{Department of Physics, University of California, San
Diego, 9500 Gilman Drive
La Jolla, California 92093}

\begin{abstract}
We observe a subharmonic gap sturucture (SGS) and the Josephson effect in superconducting scanning tunneling microscope junctions with resistances below 100 k$\Omega$. The magnitude of the $n=2$ SGS is shown to scale with the square of the junction normal state conductance, in agreement with theory. We show by analyzing the Josephson effect in these junctions that the superconducting phase dynamics are strongly affected by thermal fluctuations. We estimate the linewidth of the Josephson oscillations due to phase fluctuations, a quantity that may be important in modern theories of the subgap structure. While phase fluctuations may smear the SGS current onsets, we conclude that the sharpness of these onsets in our data is not limited by fluctuations.
\end{abstract}


\maketitle

The enhancement of the subgap tunneling current in superconducting (SC) tunnel junctions at voltages $eV=2\Delta/n$, where $\Delta$ is the SC gap and $n$ is an integer, was first observed many years ago \cite{Taylor63}. The phenomenon is termed the subharmonic gap structure (SGS) and is characterized by temperature independent excess current onsets at these voltages. Renewed interest in this effect over the last decade \cite{Maezawa94,VanDerPost94,Bratus95,Cuevas96,Averin95,Scheer97,Scheer98,Ludoph00,Scheer00,Suderow00,Cron01} was driven in part by advances in fabrication of controllable superconducting point contacts. The emergence of new systems, such as Josephson junction qubits where quasiparticle currents play a role in the decoherence of the qubit state \cite{Martinis03}, also calls for a better understanding of subgap quasiparticle tunneling processes.

Early theoretical attempts to describe this effect in ``good" (low transparency) tunnel junctions offered several possible mechanisms, most notably multiparticle tunneling \cite{Schrieffer63} (MPT) and Josephson self coupling \cite{Werthamer66}. Both mechanisms can account for the observed subharmonic structure, but their predictions for the dependence of the effect on the junction transparency is different \cite{Hasselberg74}. In the limit of high transparency contacts, Klapwijk {\it et al.} \cite{Klapwijk82} proposed multiple Andreev reflections (MAR) as a plausible mechanism. Experiments using controllable atomic break junctions \cite{VanDerPost94,Scheer97} have traced the behavior of the SGS through a wide range of junction transparencies from the tunneling to the point contact regimes. These have led to a theoretical effort \cite{Bratus95,Cuevas96,Averin95,Arnold87} to interpolate between the descriptions of SGS in the low (MPT) and high (MAR) transparency limits. 

The theories of Refs.\ \cite{Bratus95,Cuevas96,Averin95}, while successful in explaining the experimental results, formally rely on a well defined time evolution of the SC phase across the junction with the Josephson frequency $\omega=2eV/\hbar$ in the presence of a bias voltage. Indeed it is well known that the SGS in atomic junctions is smeared in the presence of rf noise \cite{Suderow00}. However, a theoretical description of this smearing within the framework of Refs.\ \cite{Bratus95,Cuevas96,Averin95}, and its connection with noise induced phase fluctuations and broadening of the Josephson oscillations is not yet established.

\begin{figure}
\centering
\includegraphics[width=\columnwidth, trim= 0 20 0 20]{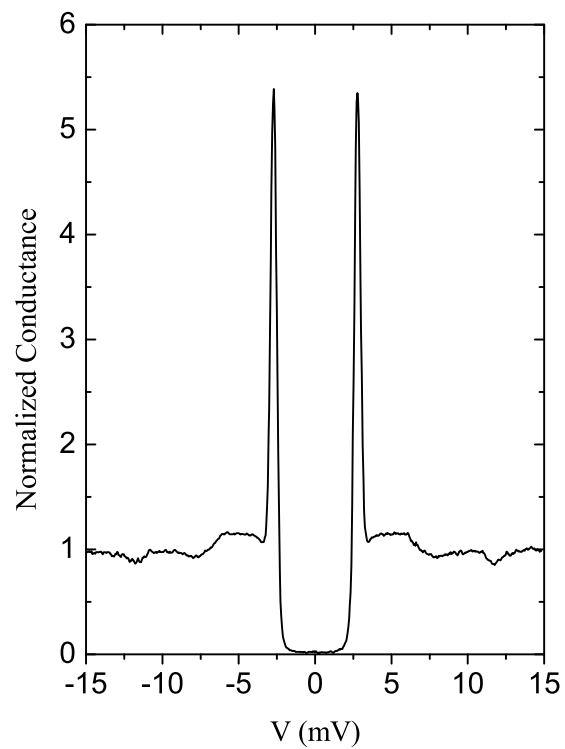}
\caption{Typical tunneling conductance for a superconducting STM vacuum tunnel junction at T=2.1 K and high junction resistance $R_N\sim 10$ M$\Omega$.}
\label{fig1} 
\end{figure}

We report here on experiments using SC scanning tunneling microscope (STM) tips to form vacuum tunnel junctions against SC films. For moderate junction transparencies \cite{Transparency} we observe the Josephson effect and the subgap structures at $eV=2\Delta/2$ and $eV=2\Delta/3$. In contrast with the break junction experiments \cite{Scheer97,Scheer98,Scheer00}, where thermal and noise induced phase fluctuations were kept minimal \cite{Goffman00} and their effect on the SGS was negligible, the noise temperature in the present work is quite high. With the observation of a smeared SGS in the current-voltage characteristics of our junctions, and a measurement of the noise temperature based on the Josephson effect, this work represents a first attempt at studying the effects of phase fluctuations on the SGS, which so far has been only qualitative. Furthermore, this work emphasizes the onset of the SGS in the low transparency tunneling regime, where most of the existing literature is concerned with high transparency near-contact junctions.

All measurements were performed at T=2.1 K in a home built STM. The scan signals, thermometer, and bias voltage lines were filtered before entering the cryostat using commercial feedthrough filters. Additional filtering of the bias leads was provided by an RC filter mounted near the junction at 2.1 K. The tunneling current is measured with a current-voltage converter (IVC) at room temperature that is connected to the junction via a coaxial cable. A resistor mounted at the input of the IVC and the capacitance of the cable act as a low pass filter above $\sim$20 MHz. Since we do not employ special microwave filters in these experiments, and the effectiveness of the filters that we use severely degrades at microwave frequencies, we expect high frequency noise from the room temperature electronics to reach the junction and contribute to an elevated noise temperature.

Superconducting STM tips were prepared as detailed in Ref.~\cite{Naaman01a} by depositing a Pb(5000 \AA)/Ag(30 \AA) proximity bilayer onto mechanically cut Pt$_{0.8}$Ir$_{0.2}$ tips. A freshly cleaved graphite crystal was placed near the tips at the time of the deposition and a Pb/Ag film was deposited on it as well, serving as the sample. We verified the quality of the tip by scanning the tip over the sample in the imaging mode. Typical spatial resolution obtained with these tips is $\sim$10 \AA. Current-voltage ($I-V$) and conductance ($dI/dV$) curves were then taken to verify the superconducting properties of the junctions. A typical $dI/dV$ curve is shown in Fig.\ \ref{fig1} where the sharp conductance peaks at voltages $V=\pm2\Delta_{\rm Pb}/e$, the small leakage conductance below the gap, and the Pb phonon structure confirm that these are good tunnel junctions.

\begin{figure}
\centering
\includegraphics[width=\columnwidth, trim= 0 0 0 20]{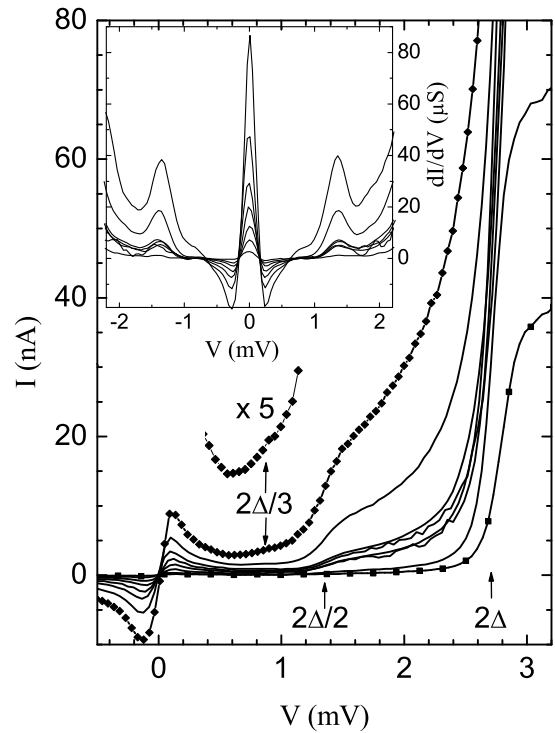}
\caption{$I-V$ characteristics for $R_N$ between 70 k$\Omega$ (squares) and 8 k$\Omega$ (diamonds) showing subharmonic gap structure and fluctuation dominated Josephson effect. The magnified ($I\times 5$) portion of the lowest resistance curve shows the $n=3$ structure more clearly. Inset: numerically smoothed and differentiated data.}
\label{fig2}
\end{figure}

A sequence of $I-V$ curves was recorded as the junction normal state resistance was lowered by moving the tip closer to the sample, thereby increasing the junction's transparency as $|T|^2\propto R_N^{-1}$. Fig.\ \ref{fig2} shows selected $I-V$ curves from such a sequence, for junction resistances ranging from 70 k$\Omega$ to 8 k$\Omega$. New features in the tunneling characteristics emerge as the junction resistance is lowered - a current peak near zero bias that is the signature of pair tunneling~\cite{Naaman01b}, and the subharmonic gap structure at voltages $2\Delta/2e$ and $2\Delta/3e$. When the data are numerically differentiated (Fig.\ \ref{fig2} inset) the subgap structure is seen as conductance peaks at V=1.35 mV and V=0.85 mV for the $n=2$ and $n=3$ subharmonics respectively. 

\begin{figure}
\centering
\includegraphics[width=\columnwidth, trim= 0 0 0 10]{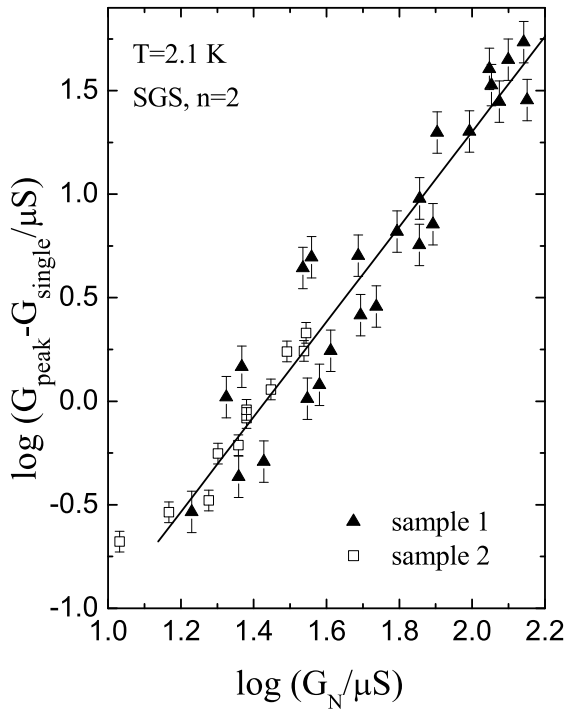}
\caption{The differential conductance at $eV=2\Delta/2$ versus the junction normal state conductance (log scale) for two different junctions (sample 1 is shown in Fig.~\ref{fig2}). The single quasiparticle background that we have subtracted was deduced from curves measured at higher junction resistances for which the normalized conductance is independent of $R_N$. The solid line represents a slope of 2.3.}
\label{fig3}
\end{figure}

In order to determine the transparency dependence of the SGS magnitude, we plot the peak conductance, after subtracting the single quasiparticle background, for the $n=2$ structure as a function of the junction normal state conductance on a logarithmic scale in Fig.~\ref{fig3}. The linear relation obtained from this graph suggests that $G_{SGS}\equiv G_{\rm Peak}-G_{\rm Single}\propto G_N^\alpha$, and from a fit we obtain $\alpha=2.3\pm 0.2$, in good agreement with the expected $|T|^4\sim G_N^2$ dependence. We observe a similar dependence in all junctions measured to date, with the exponent $\alpha$ ranging from 1.8 to 2.3. Strictly speaking, it is the current jump at this voltage that is expected to scale with $|T|^4$, but since the width of this onset in our data is independent of $G_N$, the behavior of the differential conductance peak reflects that of the current jump. Unfortunately, the present data do not allow us to determine the transparency dependence of the $n=3$ feature, although its magnitude relative to the $n=2$ feature is consistent with the theoretically predicted $|T|^6\sim G_N^3$ dependence.

Were the junctions single channel contacts, their transparency would have saturated to 1 at $R_N=12.9$ k$\Omega$. The fact that we still observe tunneling-like $I-V$'s down to 8 \nolinebreak k$\Omega$ suggest that our tips accomodate a few channels. This conclusion is also consistent with the $\sim$10 {\AA} spatial resolution that we estimate from working in the imaging mode. The relatively large scatter in the data in Fig.\ \ref{fig3} can thus be understood as resulting from atomic rearrangements in the tip during the time of the experiment. Such rearrangements may change the transparencies of the individual channels in the junction; while there are many different combinations of channel transparancies that can result in the same normal state conductance, the $I-V$ characteristics below the SC gap are highly nonlinear and therefore sensitive to the particular channel content of the junction \cite{Scheer97}.  
 
\begin{figure} 
\centering
\includegraphics[width=\columnwidth, trim= 0 0 0 10]{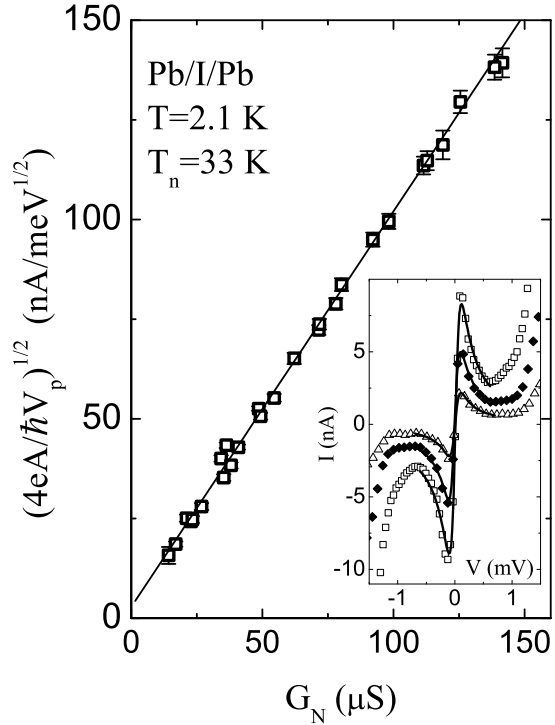}
\caption{A plot of $\sqrt{(4e/\hbar)A/V_p}$ vs. the junction normal state conductance. Inset: fits of the $I-V$ characteristics using the phase diffusion model of Ref.\ \cite{Ivanchenko69}.}
\label{fig4}
\end{figure}

In light of the theoretical descriptions of the SGS in Refs.\ \cite{Bratus95,Cuevas96,Averin95} it is important to further characterize the phase dynamics of these junctions. The evolution of the Josephson current as a function of the junction normal state resistance allows us to quantify the strength of the phase fluctuations in the junction. Because of the junctions' small size and the high resistance associated with it, the Josephson binding energy $E_J=\pi\hbar\Delta/4e^2R_N$, which is the energy scale for the coupling of the phases across the junction, is comparable to $k_BT$. In this regime, where $E_J\approx k_BT$, the motion of the phase difference is diffusive \cite{Ivanchenko69,Ambegaokar69}, and the pair current is therefore associated with finite voltages. The $I-V$ characteristics due to pair tunneling for small $E_J/k_BT$ were shown by Ivanchenko and Zil'berman \cite{Ivanchenko69} to have the form $I(V)=A\times V/(V^2+V_p^2)$, where $A=I_c^2Z_{\rm env}/2$, $I_c=2eE_J/\hbar$, $V_p=(2e/\hbar)Z_{\rm env}k_BT_n$, $Z_{\rm env}$ is the impedance of the junction's environment, and $T_n$ is the effective noise temperature. We fit the $I-V$ curves using the equation above with $A$ and $V_p$ the only fitting parameters (Fig.\ \ref{fig4} inset). A plot of $\sqrt{(4e/\hbar)A/V_p}$ as a function of $G_N$ yields a linear graph with a slope of $I_cR_N/\sqrt{k_BT_n}$ \cite{Naaman01b}. Such a plot (Fig.\ \ref{fig4}) is constructed from analysis of the data of Fig.\ \ref{fig2}. From the slope of the graph, and using the Ambegaokar-Baratoff relation \cite{Ambegaokar63} to find $I_cR_N=1.67$ mV in this junction, we estimate the noise temperature $T_n=33\pm1$ K \cite{SteinbachNote}, and the environment impedance $Z_{\rm env}=78\pm7 \Omega$ close to the impedance of free space as expected \cite{SETbook}.

The quantity of importance to the dicussion here is the linewidth of Josephson oscillations around $\dot{\phi} =2eV/\hbar$. The mechanism for the formation of the SGS within the theory of Bratus' {\it et al.} \cite{Bratus95} is the inelastic Andreev scattering of quasiparticles impinging on the junction into $n$ sidebands. The splitting in the spectrum of scattered waves into sidebands is induced by the time varying phase across the junction with the Josephson frequency $\omega=2eV/\hbar$. Our experiments show that in the presence of thermal phase fluctuations there is a considerable broadening in the time evolution of the phase \cite{Naaman01b}, where the linewidth $\Gamma$ around the Josephson frequency depends on the effective noise temperature and the impedance of the junction's environment. In the RSJ model this broadening can be estimated \cite{Likharev} as $\Gamma\approx(2e/\hbar)^2Z_{\rm env}k_BT_n=2eV_p/\hbar$. For the data shown here $V_p=0.11\pm 0.01$ mV, resulting in $\Gamma\sim 300$ GHz. 

\begin{figure}
\centering
\includegraphics[width=\columnwidth, trim= 0 20 0 0]{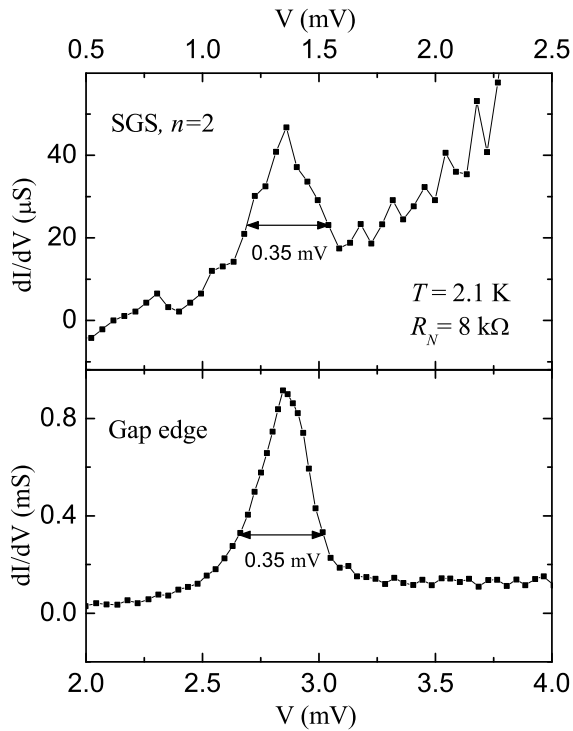}
\caption{Compaison of the widths of the $n=2$ SGS (top panel) and the gap edge (bottom panel). Note the different conductance scale, and the offset voltage scales.}
\label{fig5}
\end{figure}

We expect that the effects of phase fluctuations would be to broaden the subgap features~\cite{Suderow00,Arnold87} similarly to the broadening of microwave induced Shapiro steps \cite{Naaman01b}. It is therefore important to determine whether the broadening of the SGS features in our junctions (Fig.~\ref{fig2} inset), with full width of approximately 0.35~mV, arise from phase fluctuations. Since any subgap quasiparticle tunneling process would eventually involve the injection of a quasiparticle at the gap edge of the SC electrode, the SGS current onsets can be no sharper than the current rise at $eV=2\Delta$. A~well documented phenomenon in Pb, of which our junctions are made, is the apparent broadening of the gap edge in tunneling experiments~\cite{Giaver62,Rochlin67,Lykken71}. This broadening is intrinsic in bulk Pb, and is a result of the $\sim$10\% anisotropy in the Pb order parameter~\cite{Blackford69,Short00}. Figure~\ref{fig5} shows a comparison between the conductance peaks at the gap edge and at the $n=2$ SGS. The fact that the widths of these peaks are approximately the same, and larger than the characteristic broadening of the Josephson oscillations $(\hbar/2e)\Gamma$, suggests that the sharpness of the SGS in our junctions is not fluctuation limited, but rather it is intrinsic to the superconductors comprising the junction. We would expect that with increasing strength of the phase fluctuations, the width of the SGS would eventually increase above the width of the gap edge.

To summarize, we have observed the $n=2$ and $n=3$ subharmonic gap structure in superconducting STM vacuum junctions, and found a $G_N^2$ dependence for the magnitude of the $n=2$ SGS. This experiment has been done under conditions where strong phase fluctuations are present, and by analyzing the Josephson effect in these junctions we have determined the broadening of the Josephson oscillation linewidth due to these fluctuations. Since modern theories for the origins of the excess subgap current~\cite{Bratus95,Cuevas96,Averin95,Arnold87} suggest that the subharmonic gap structure and the ac Josephson effect are intimately related, it could be expected that strong phase fluctuations will alter the shape of the SGS. Nevertheless, we conclude that the apparent smearing of the observed SGS in our data is most likely due to the intrinsic gap anisotropy in Pb, rather than phase fluctuations. Experiments such as described here, could in principle, provide a testing ground for theoretical descriptions of the effects of fluctuations on the SGS. The complications introduced by the intrinsic width of the gap edge in Pb, can be easily overcome by fabricating the junctions from Pb$_{1-x}$Bi$_x$ alloys with $x\sim0.10-0.15$ in which the mean free path is shorter, and as a result, the gap edge is sharper, with characteristic widths at 2~K of 0.068~meV and 0.048~meV, for $x=0.10~{\rm and}~0.15$ respectively~\cite{Campbell66,Dynes78}. 

This experiment demonstrates that STM studies with well characterized SC tips are suitable for careful measurements of the subgap structure. Additional advantage of STM experiments is that in contrast with break junctions that are necessarily symmetric, the STM setup offers the possibility to study asymmetric junctions (e.g. Pb/I/NbSe$_2$) \cite{Naaman03}. We note that the STM has been used before \cite{Scheer98,Suderow00} to observe the SGS, but unfortunately a quantitative analysis of the phase fluctuations in those experiments was not reported. We argue that the measurement of the phase fluctuations rather than their suppression may further help to test the existing theories. 

We would like to thank L. Bokacheva, W. Teizer, and the UCSD
Physics Electronics Shop. We have benefitted from discussions with H.F. Hess, H. Suderow, E. Bascones, and V. Shumeiko. This work was supported by DOE Grant number DE-FG03-00ER45853.


\begin{thebibliography}{00}
\bibitem{Transparency} We shall use the term transparency to mean $|T|^2$, where $|T|$ is the tunneling matrix element.
\end{thebibliography}
\end{document}